\documentclass[12pt]{article}
\usepackage{hyperref}
\usepackage{epsf}

\usepackage{amssymb}

\def\a{\alpha}
\def\b{\beta}
\def\g{\gamma}
\def\G{\Gamma}

\def\d{\delta}
\def\e{\varepsilon}

\def\k{\kappa}

\def\beq{\begin{equation}}
\def\eeq{\end{equation}}
\def\beqn{\begin{eqnarray}}
\def\eeqn{\end{eqnarray}}
\def\ba{\begin{eqnarray}}
\def\ea{\end{eqnarray}}

\def\m{{\tt -}}

\def\l{\langle}

\def\xprim2bar{\overline{x}^{\prime\prime}}

\def\beq{\begin{equation}}
\def\eeq{\end{equation}}

\newcommand{\beqa}{\begin{eqnarray}}
\newcommand{\eeqa}{\end{eqnarray}}

\let\a=\alpha   \let\b=\beta   \let\g=\gamma   \let\d=\delta
\let\e=\epsilon         
    \let\k=\kappa  \let\l=\lambda  \let\m=\mu

\let\G=\Gamma     \let\L=\Lambda

\newcommand{\eq}[1]{Eq.~(\ref{#1})}

\def\g0p{g_0^\prime}

\let\a=\alpha   \let\b=\beta   \let\g=\gamma   \let\d=\delta
\let\e=\epsilon         
    \let\k=\kappa  \let\l=\lambda  \let\m=\mu

\let\G=\Gamma     \let\L=\Lambda

\newcommand{\be}{\begin{equation}}
\newcommand{\ee}{\end{equation}}
\newcommand{\bea}{\begin{eqnarray}}
\newcommand{\eea}{\end{eqnarray}}

\usepackage{graphics}
\begin{document}
\begin{titlepage}
\setcounter{page}{1}
\vspace{2.0cm}
\begin{center}{\Large \bf On the construction of Chern--Simons terms in the
presence of flux}
\vspace{.17in}

\vspace{.6cm}

       {\large Nikos Irges and Mirian Tsulaia}

\vspace{.12in}

\vspace{.5cm}

~\\
{\it
Department of Physics, University of Crete, 71003 Heraklion, Greece\\}
~\\
\end{center}
\vspace{4cm}

\begin{abstract}

We develop a method for relating the
boundary effective action associated with
an orbifold of the D+1--dimensional theory of a
p--form field to D--dimensional fluxed Chern--Simons
type of terms. We apply the construction to derive
from twelve dimensions the
Chern--Simons terms of the eleven dimensional
supergravity theory in the presence of flux.

\end{abstract}

\smallskip
\end{titlepage}
\renewcommand{\theequation}{\arabic{section}.\arabic{equation}}

\setcounter{equation}0

\section{Introduction}

Compactifications of the superstring theories
to lower dimensions in the presence of nontrivial background R--R and NS--NS
fluxes
are the subject of active current research interest.
Motivated first from being a possible solution to the hierarchy problem
\cite{Giddings:2001yu},
they revealed a rich theoretical and phenomenological structure
in compactified $D=4$ low energy effective field theories.
In particular, flux compactifications offer the possibility of
lifting all moduli in lower dimensions and possess semi--realistic cosmology in
the presence
of a positive cosmological constant \cite{Kachru:2003aw,Kachru:2003sx}.
The main effort in flux compactifications
has been concentrated earlier on various compactifications of
M--theory \cite{Lukas:1998yy},
more recently on compactifications of
type IIB superstring theory on six dimensional (generalized) Calabi--Yau
manifolds and toroidal orientifolds
(for a review see \cite{Grana:2005jc})
but much less is known about type IIA
superstring flux compactifications.
Recently, the compactification of type IIA supergravity
to four dimensions where the compact space is a Calabi--Yau manifold
or a $T^6/Z_3$ orientifold was studied in detail
\cite{Kachru:2004jr,DeWolfe:2005uu}.
There it was shown that in these models all geometric moduli can be stabilized
classically.
The latter results give the strong hint that flux compactifications of
the type IIA theory can be at least as interesting as the ones of type IIB.

In this context, it seems particularly interesting
to address the issue of the construction of the Chern--Simons
terms of type IIA supergravity in the presence of background flux.
Let us recall that in the absence of flux
the Chern--Simons terms $H_3\wedge C_3 \wedge F_4 $ and
$-B_2\wedge F_4 \wedge F_4$ (we use the notation of \cite{Polchinski:1998rr})
are equivalent, since the total derivative vanishes on the boundary.
However, this is not the case in the presence of flux
and therefore the definition of Chern--Simons
terms becomes more subtle \cite{Kachru:2004jr,DeWolfe:2005uu}.
This can be seen by noticing that Chern-Simons terms in the presence
of (topological) flux are, in general, not invariant
under (large) gauge transformations.
The correct definition of a Chern--Simons term $\Gamma$ on a $D$ dimensional
manifold $M_D$ in the presence of topological flux is
\be
{\int_{M_D} \Gamma } = {\int_{M_{D+1}} d\, \Gamma },
\ee
where the $D$--dimensional  manifold $M_D$ is a boundary of a $D+1$
dimensional manifold $M_{D+1}$.
However this is not enough in order to resolve the ambiguity.
Only if the (by construction) gauge invariant right hand side can be
reduced onto the boundary in a smooth, gauge invariant way,
then one can obtain the most general $\G$,
consistent with $D$--dimensional gauge invariance.
In this way one can construct a general class of Chern--Simons terms
(i.e. excluding the terms containing curvature forms
which are present in the type IIA theory
\cite{Diaconescu:2000wy,Diaconescu:2003bm})
relevant for flux compactifications.

\section{The orbifold construction}

Here we will present a method of obtaining the type of Chern--Simons terms
discussed above by employing the $U(N)$ orbifold gauge theory construction of
\cite{Irges:2004gy}.
The idea is to encode the non--trivial topology and any symmetry breaking
occurring
at the orbifold fixed point into the transition functions associated with
the gauge fields living on those charts that contain the orbifold
fixed point in their intersection.
Then, shrinking the intersection to a point (the fixed point),
if the value of the transition function at that point is non--zero,
the boundary effective theory will develop contributions
that look like flux effects along the boundary.
Even though for concreteness we will discuss only
the case of 11--dimensional supergravity, we believe that
the method can be in general applied
to the problem of construction of gauge invariant,
fluxed Chern--Simons terms in any dimension.

Let us therefore consider $11$--dimensional supergravity
and try to find its most general flux extension.
To do so, we will follow \cite{Witten:1996md} and \cite{DeWolfe:2005uu}
and introduce flux in the $11$--dimensional action via a
$12$--dimensional manifold with boundary.
We start by assuming the existence of a 12--dimensional manifold
${\cal M}_{12}$ on which a $3$--form field can be locally defined.
The coordinates on this manifold are $z^{M}=(x^{\m}, x^{11})$.
The 11--dimensional coordinates are $x^{\m} = (x^{\hat \m}, x^{10})$.
We would like then to construct the ${Z}_2$ orbifold
of this theory by projecting out
by the reflection ${\cal R}$, which acts on the 12-dimensional
coordinates as
\be
{\cal R}z = {\overline z},\hskip 1cm {\overline z} = (x^{\m}, -x^{11}).
\ee
The coordinate $x^{11}$ can be either space--like or time--like.
The action of ${\cal R}$ on a rank--$r$ tensor field $C(z)$ is defined as
\be
\left({\cal R}\,C_{M_1M_2 \cdots M_r}\right)(z) =
\a_{M_1}\a_{M_2}\cdots \a_{M_r}\,C_{M_1M_2 \cdots M_r}({\cal R}\,z) \,,
\label{Rtrans}
\ee
where no sum on the $M_i$ is implied on the right hand side.
The intrinsic parities are defined by $\a_{\m}=1$ and $\a_{11}=-1$.
Parity of the exterior derivative of forms can be easily derived using that
$[{\cal R}\,,\partial_M] =  0$.
At the fixed point of the orbifold, $x^{11} = 0$, an
$11$-dimensional theory living on the boundary manifold ${\cal M}_{11}$
can be consistently defined.
We would like this theory to be
somehow related to the $11$--dimensional supergravity of \cite{Cremmer:1978km}.

We compactify $x^{11}$ on a circle of radius $R$.
\footnote{By taking the radius of the circle to infinity
one can describe in this way the non-compact version of the orbifold.}
The gauge invariant construction of the orbifold proceeds by
defining separate $3$--form $U(1)$ gauge fields on overlapping
charts that provide an open cover for the $12$--dimensional space.
The minimum number of such overlapping open sets
in the $x^{11} = 0$ neighborhood is two,
let us call them $O^{(+)}$ and $O^{(-)}$ and their overlap
$O^{(+-)}=O^{(+)} \cap \, O^{(-)}$.
On each open set there is a $3$--form gauge field that
under a $12$--dimensional $U(1)$ gauge transformation
transforms with its own $2$--form gauge function
\begin{eqnarray}
{\rm on}\; O^{(\pm)}:\hskip.5cm \d\; C^{(\pm)} = d\, \L^{(\pm)}.
\end{eqnarray}
One requires that the $3$--forms on $O^{(+-)}$ (where they are both defined)
are related by a gauge transformation:
\begin{eqnarray}
C^{(+)} = C^{(-)} + d\, g^{(+-)},\hskip 1cm C^{(-)} = C^{(+)} + d\, g^{(-+)}\,.
\label{rel2}
\end{eqnarray}
The $U(1)$--valued $2$--forms $g^{(+-)}$ and $g^{(-+)}$ are
transition functions and they are defined on the overlap of charts $O^{(+-)}$.
The gauge invariant $4$--form field strength
\be
G = d\, C^{(+)} = d\, C^{(-)},\hskip 1cm \d\, G = 0,
\ee
does not depend on the chart label since $d^2 = 0$
and therefore it is uniquely defined throughout ${\cal M}_{12}$.
The Bianchi identity $d\, G = 0$
asserts that $G$ is closed and so defines an element of
$H^4({\cal M}_{12}, R)$, the fourth cohomology class of
${\cal M}_{12}$ with real coefficients.

The consistency of the system of equations
\eq{rel2}  requires the existence of a 1-form $\chi$ defined on $O^{(+-)}$ such
that
\begin{equation}
g^{(+-)} + g^{(-+)} = d\, \chi.\label{gluingforms}
\end{equation}
Furthermore, the form of \eq{rel2} is preserved under the gauge transformations
\begin{eqnarray}
&& \d\, g^{(+-)} = \L^{(+)} -\L^{(-)} + \frac{1}{2}\, d\, (\l^{(+-)} +
\l^{(-+)})\nonumber \\
&& \d\, g^{(-+)} = \L^{(-)} -\L^{(+)} + \frac{1}{2}\, d\, (\l^{(+-)} -
\l^{(-+)}),  \label{Ggt}
\end{eqnarray}
for some $1$--forms $\l^{(+-)}$ and $\l^{(-+)}$.
Combining \eq{gluingforms} with \eq{Ggt} we derive the gauge
transformation of $\chi$
\be
\d\, \chi = \l^{(+-)} + d\, \phi_0
\ee
for some 0-form $\phi_0$.
The above is essentially an appropriate generalization
of a 1-form fibre bundle construction to higher order forms,
known as gerbes, see for example \cite{Lupercio:2004xc}.   

In order to introduce flux in the boundary theory,
one must construct a non-dynamical $4$--form on the overlap $O^{(+-)}$, let us
call it ${\overline G}$, so that in the limit where the overlap shrinks to
a single point (the orbifold fixed point) it goes to an
$x^\m$-dependent function, which is in general not zero
(examples of analogous constructions in Yang--Mills theories have been
constructed in
\cite{Irges:2004gy}).
If in addition this $4$-form obeys
\be
d\, {\overline G} =0, \hskip 1cm \d\, {\overline G} =0,
\ee
then it can be safely added to the gauge invariant field strength $G$.
Its effect at the orbifold fixed point will be a non-zero flux on ${\cal
M}_{11}$.
One such $4$-form is
\be
{\overline G} = {\overline g} \wedge {\overline g}, \hskip 1cm
{\overline g} = g^{(+-)} + g^{(-+)}
\ee
provided that $\l^{(+-)}$ is exact, i.e. $\l^{(+-)} = d\, l_0$, that is
provided that ${\overline g}$ is gauge invariant.
Since ${\overline G}= d\, {\overline C}$, with ${\overline C}=\chi\wedge d\chi$,
for a gauge invariant ${\overline g}$ under a gauge transformation
we have $\d {\overline C} = d((l_0 + \phi_0)\wedge d\chi)$.

Next, we have to define the action of the reflection operator
on the geometry and the fields.
For concreteness take
\begin{eqnarray}
O^{(+)} \,=\, (-\e ,\pi R + \e) & \mbox{and} &
O^{(-)} \,=\, (-\pi R - \e ,\e)
\end{eqnarray}
with overlap $O^{(+-)} = (-\e ,\e )$, where $0< \e < + \pi R/2$
(we concentrate on the neighborhood of $x^{11}=0$ only).
The reflection operator maps ${\cal R}\,O^{(\pm)} \,=\, O^{(\mp)}\,$,
${\cal R}\,O^{(+-)} \,=\, O^{(+-)}$.
The transformation ${\cal R}$ can be defined also to act on tensor fields
defined
on $O^{(\pm)}$ giving as result tensor fields defined on $O^{(\mp)}$.
On the overlap, we define
\be
{\cal R}\, C^{(+)} = C^{(-)}.
\ee
We are interested in the action of ${\cal R}$ on the fields ${\overline g},
\chi,\l^{(+-)} $
and $\phi_0$.
It is not hard to check that one can consistently define
\begin{eqnarray}
&& {\cal R}\, {\overline g} = {\overline g}, \hskip 2.5cm
{\cal R}\, {\chi} = {\chi} \nonumber \\
&& {\cal R}\, \l^{(+-)} = \l^{(+-)}, \hskip 1cm
{\cal R}\, \phi_0 = \phi_0.
\end{eqnarray}
Furthermore, if we choose  ${\overline g}$ to be gauge invariant
then
\be
{\cal R}\, l_0 = l_0.
\ee
As far as the projections that the above actions imply,
it is only the components of the $3$-form field and its
field strength along the boundary that survive.

The boundary theory is now simple to obtain.
One is instructed to construct
all possible gauge invariant terms using the
original $12$-dimensional fields, $G$ and ${\overline G}$ in our case
and then take the limit where the overlap $O^{(+-)}$ shrinks
to the fixed point of the orbifold action.
The limit $\e\longrightarrow 0$ can be taken with
the only essential ingredient needed being that
\be
\lim _{\e \longrightarrow 0} {\overline G} \equiv G^{flux}(x^\m) \ne 0
\ee
and a similar condition on the limit of ${\overline C}$
(in a way that quantum effects do not trigger the appearance
of new dynamical fields on the boundary).
The result is simple for the terms involving $G$.
The $12$--dimensional interaction term is
\be
S_{12}^{int} = -\frac{1}{2\k_{12}^2}
\int_{{\cal M}_{12}} (G+\a {\overline G})\wedge (G+\b {\overline G})\wedge (G+\g
{\overline G}),\label{12CS}
\ee
where $\a$, $\b$ and $\g$ are, to this end, arbitrary coefficients
in the absence of a symmetry that can relate them.
On the boundary, it reduces to the Chern--Simons interactions
\begin{eqnarray}
S_{11}^{CS} = &-&\frac{1}{12\k_{11}^2} \int_{{\cal M}_{11}}
\Bigl\{ {\hat C} \wedge {\hat G} \wedge {\hat G} \nonumber\\
&+& \a^\prime \, {\hat C} \wedge {\hat G} \wedge {G^{flux}} +
\b^\prime \, {\hat C} \wedge {G^{flux}} \wedge {G^{flux}} +
\g^\prime \, {C^{flux}} \wedge {G^{flux}} \wedge {G^{flux}}
\Bigr\},\nonumber\\
\label{11CS}
\end{eqnarray}
where ${\hat C}$ and ${\hat G}$ are essentially
the $12$--dimensional $3$--form and $4$--form
evaluated at $x^{11}=0$.

Let us emphasize that the procedure described above does not determine
the coefficients $\a^\prime$, $\b^\prime$ and $\g^\prime$.
In order to obtain their precise values,
either a direct computation of boundary counterterms
or the implementation of some extra symmetry is needed.
Regarding the first option, given that in a field theory
boundary counterterms appear beyond the classical level
\cite{Sym} and their computation requires
a good control of the theory at least at the perturbative level,
it seems out of reach at present for any 12--dimensional theory.
This difficulty suggests one to consider the possibility of
supersymmetrizing the theory since in supersymmetric
theories it is not uncommon that coefficients that appear
in the effective action at the quantum level are completely
fixed by means of symmetry.
In the case of eleven dimensional supergravity these coefficients
can be fixed from knowing that the full Lagrangian
which, besides \eq{11CS},
includes also an Einstein--Hilbert action and a kinetic term for the three
form, must be
a bosonic part of an effective action describing the
eleven dimensional  supergravity in the presence of fluxes.
The quantity $C^{flux}$ can be now interpreted as a solution of the bosonic
equations of motion of the eleven dimensional supergravity
 while ${\hat C}$ is a quantum
fluctuation around it.
In this interpretation the flux part of the effective action in
general  can depend on the coordinates of $M_{11}$
and can break a part of supersymmetries.
Requiring agreement with the Chern--Simons
term obtained in \cite{DeWolfe:2005uu} fixes
\be
\g^\prime \, =\, 0,
\ee
as well as
\be
\a^\prime = \b^\prime =3.
\ee
The choice of $\gamma^\prime$ is  natural
since it means that one omits  a term in the effective action which corresponds
to
a non--zero vacuum energy.
The coefficients $\a^\prime$ and $\b^\prime$
(the value of $G^{flux}$)
can be alternatively
fixed  by compactifying the theory on
$S1\times CY_3$ and matching the resulting four dimensional theory
to a specific ${\cal N}=2$  gauged supergravity theory
\cite{Kachru:2004jr,DeWolfe:2005uu}.

It would be nice though if it was possible to fix
such coefficients directly via the higher
dimensional theory imposing e.g. appropriate supersymmetry transformations.
Unfortunately, it is difficult to say how the explicit realization
of supersymmetry in dimensions higher than eleven (see \cite{Rudychev:1997ui}
for the review on higher dimensional superalgebras)
works.  It is also known that in dimensions higher than eleven
one necessarily obtains fields of spin larger than two,
for which no analogue of the Einstein--Hilbert action is known.
However let us note that the eleven dimensional superalgebra
\be
\{ Q_\alpha Q_\beta \} = P^\mu \g_{\alpha \beta} + \gamma^{\mu \nu}_{\alpha
\beta}Z_{ \mu \nu} +
\gamma_{\alpha \beta}^{\mu_1 \cdots \mu_5} Z_{\mu_1 \cdots \mu_5}
\ee
can be embedded in the 12--dimensional $ {\cal N}=1$ superalgebra with signature
$(10,2)$ in a $Z_2$ parity invariant way. Alternatively
it can also be obtained by orbifolding
the 12--dimensional $ {\cal N}=2$ superalgebra
\begin{equation}
\{ Q^i_\alpha Q^j_\beta \} = ( \tau_a)^{ij}(\gamma^{MN}_{\alpha \beta}Z^a_{MN} +
\gamma_{\alpha \beta}^{M_1 \cdots M_6} Z^{a+}_{M_1 \cdots M_6})
+\epsilon^{ij}( C_{\alpha \beta}Z +
\gamma_{\alpha \beta}^{M_1 \cdots M_4} Z_{M_1 \cdots M_4}),
\end{equation}
where $i=1,2$ and $a=1,2,3$.
Note also that the supermultiplet which is a
representation of these twelve dimensional algebras
does contain a three form field
\cite{Bars:1987nr}, which is expected to reduce to the eleven
dimensional three form after the orbifold projection along the lines we have
described. It is therefore not unlikely that the whole action obtained
by supersymmetrizing Eq. (\ref{12CS}) reduces to some generalization
of the 11D-supergravity action upon orbifolding.
Finally, a possible mathematical handle on the nature of the 12--dimensional multiplet
could be the fact that it is the lowest lying
Euler multiplet \cite{Ramond} associated with the symmetric space
$E_6/(SO(10)\times SO(2))$ \cite{Boya}.

\section{Conclusion}

We presented a possible method for constructing gauge invariant flux
extensions of Chern--Simons terms in $D$--dimensions via an orbifold
construction.
This can be achieved by formulating the theory on
a manifold of dimension $D+1$ of which the original
manifold is a boundary, as suggested in \cite{Witten:1996md}.
Gauge invariant $D+1$--dimensional
fields and gauge transformation functions can be smoothly pulled
back onto the boundary, defining
a theory automatically gauge invariant in the $D$--dimensional sense.
Remnants of a certain class of bulk gauge
transformation (transition) functions are seen as flux along the boundary.

\bigskip

\centerline{\bf Acknowledgements}

We would like to thank C. Bachas, C. Coriano and F. Knechtli for their useful comments
on the manuscript, S. P. de Alwis and  A.~K.~Kashani-Poor for discussions
and  D. Sorokin for correspondence, comments on the manuscript
and  discussions.
M. T. is supported through the European contract
MRTN-CT-2004-512194.




\end{document}